\documentclass[12pt]{article}
\usepackage{amsmath}
\usepackage{amssymb}
\usepackage{amstext}
\usepackage{amsfonts}
\usepackage{graphicx,epsfig}
\usepackage{slashbox}
\usepackage{indentfirst}
\begin{document}

\baselineskip 20pt

\title{Determining $1^{--}$ Heavy Hybrid Masses via QCD Sum Rules}
\author{Cong-Feng Qiao$^{a,b}$, Liang Tang$^{a}$, Gang Hao$^{b}$ and
Xue-Qian Li$^{a}$\\[0.5cm]
{\small $a)$ School of Physics, Nankai University, 300071, Tianjin,
China}\\
\small $b)$ Department of Physics, Graduate University, the Chinese
Academy of Sciences \\ \small YuQuan Road 19A, 100049, Beijing,
China}
\date{}
\maketitle

\begin{center}
\begin{minipage}{11cm}
The masses of $1^{--}$ charmonium and bottomonium hybrids are
evaluated in terms of QCD sum rules. We find that the ground state
hybrid in charm sector lies in $m_{H_c}=4.12\sim 4.79$ GeV, while in
bottom sector the hybrid may situated in $m_{H_b} = 10.24\sim 11.15$
GeV. Since the numerical result on charmonium hybrid mass is not
compatible with the charmonium spectra, including structures newly
observed in experiment, we tempt to conclude that such a hybrid does
not purely exist, but rather as an admixture with other states, like
glueball and regular quarkonium, in experimental observation.
However, our result on bottomonium hybrid coincide with the ``exotic
structure" recently observed at BELLE.
\end{minipage}
\end{center}

\section{Introduction}

Quantum chromodynamics (QCD) is believed to be the underlying theory
for strong interaction. It is commonly believed that the mechanism
responsible for the hadronic properties is subject to the
non-perturbative aspect of QCD. Unlike the perturbative QCD which is
well understood in some sense, we do not have a reliable and
effective way to tackle with the non-perturbative QCD effect yet. In
this respect, to get a deep insight in the physics associated with
the non-perturbative QCD is one of the most important tasks for the
society of high energy physics. So far and in this aim, to evaluate
the physical quantities of hadrons, such as hadron spectra, hadronic
transition matrix elements, parton distributions and fragmentation
functions, in most cases people have to invoke to typical
phenomenology models.

Apart from the so-called regular hadrons, the meson and baryon, QCD
also does not exclude exotic hadronic structures, like hybrid,
glueball and multiquark structures. Normally, the hybrid meson
refers to a state which contains a pair of constituent quarks and a
dynamic gluon. In the color-flux tube model, instead, the hybrid
corresponds to the structure where the gluonic degree of freedom is
excited. Even though these two pictures look different, they may be
just two sides of the same object. Thus, whether they are
reconcilable with each other or not is an interesting question.
Indeed, a careful study may shed light on the hybrid structure and
help us to get a better understanding of the non-perturbative QCD
effects.

Up to now, a many of effective methods was proposed
\cite{Buchmuller,Isgur,Swanson,Allen} in evaluating the hybrid mass
spectrum, for instance, by studying the quarkonium hadronic
transition via multipole expansion \cite{Kuang,Ke}, the bag model
\cite{T. Bar}, the flux-tube model \cite{N. Isg}, lattice QCD
\cite{C. Mic} and QCD Sum Rules \cite{I. I. Bal, J. I. Lat, H. Y.
Jin, T. Hua 1,F. K. Guo, T. Hua 2, J. Gov 2, S. L. Zhu, L. S. Kis}.
Whereas, those theoretical evaluation results diverse greatly with
each other, and hence it is hard to pin down any exotic structures
as hybrids in experiment. Therefore, further theoretical
investigations are necessary, and they were partially done. For
instance, the light hybrid masses evaluated in \cite{J. Gov 1} were
updated by Narison and his collaborators
\cite{Narison:1989aq,Narison:2009vj,Chetyrkin:2000tj} in the
framework of QCD Sum Rules. Moreover, as indicated in Refs.
\cite{Close,Hexg}, the hybrid states may not exist independently,
but may rather admixtures of hybrids with regular quarkonia or even
glueballs with the same quantum numbers. To clarify the messy
picture, a more accurate evaluation of pure hybrid spectrum is still
vital, namely one can then confront the theoretical predictions with
the experimental data to determine the hybrid component of a
hadronic state.

Among those theoretical methods in dealing with the non-perturbative
effects, QCD Sum Rules innovated by Shifman {\it et al}.\cite{M. A.
Shi} turns out to be a remarkably successful and powerful technique
for the computation of hadronic properties. By virtue of QCD Sum
Rules, hybrids with various quantum numbers and the flavors have
been investigated. For light-quark hybrids, in order to avoid the
mixing between hybrids and ordinary mesons, Ref.\cite{I. I. Bal}
considered specifically the hybrids possessing exotic quantum
numbers $1^{-+}$, and Ref.\cite{J. I. Lat} took another exotic
quantum numbers $0^{--}$ into consideration, and obtained the
relative masses and decay amplitudes. Employing the heavy quark
effective theory (HQET), Huang, Jin and Zhang evaluated the masses
and decay widths of several typical heavy-light hybrids \cite{T. Hua
2}. The heavy-quark hybrids masses were also evaluated in Refs.
\cite{J. Gov 2, S. L. Zhu, L. S. Kis} through QCD Sum Rules.

Recently, a hadronic structure with mass $4324\pm24$ MeV and width
$172\pm33$ MeV, has been observed by the BABAR Collaboration in the
$\psi(2S) \pi^+ \pi^-$ invariant-mass spectrum \cite{BAB 1}. This
structure is obviously different from $Y(4260)$ reported in
Ref.\cite{BAB 2}. Later on, two resonant-like structures are
observed in also the $\psi(2S) \pi^+ \pi^-$ mode by the Belle
Collaboration, one resides at $4361\pm9\pm9$ MeV with a width of
$74\pm15\pm10$ MeV, which coincides with what reported by the BABAR
collaboration, and another is at $4664\pm11\pm5$ MeV with a width of
$48\pm15\pm3$ MeV \cite{X. L. Wan}. A variety of theoretical
speculations has been put forward for these hadronic structures. For
instance, supposing Y(4664) still be a normal member of the
charmonium family, it was interpreted as different charmonium
states, like a $5^3S_1$ state, a $6^3S_1$ state, or even a
$5^3S_1-4^3D_1$ mixed state \cite{G. J. Din}. The $Y(4664)$ was also
interpreted as a baryonium state \cite{C. F. Qia}, the radial
excited state of
$\frac{1}{\sqrt{2}}(|\Lambda_c\bar{\Lambda}_c\rangle+|\Sigma_c^0
\bar{\Sigma}_c^0\rangle)$, and a $f_0(980)\psi^\prime$ molecule
\cite{Guo:2008zg}. Starting from the QCD Sum Rules, Albuquerque {\it
et al}. \cite{R. M. Alb} computed the mass of $Y(4664)$ based upon
the assumption that it is a vector $cs\bar{c}\bar{s}$ tetraquark
state. In the literature, for convenience, $Y(4361)$ and $Y(4664)$
are usually named as $Y(4360)$ and $Y(4660)$, respectively.

Since a series of newly observed ``exotic" states in charmonium
energy region is $1^{--}$ hadrons, in this paper we reinvestigate
the $1^{--}$ charmonium hybrid, which is constructed by a pair of
charm-anticharm quarks and a gluon, by means of the QCD Sum Rules.
In our calculation, the interpolating current of the charmonium
hybrid is chosen to be
$g_s\bar{\psi}\gamma^\nu\gamma_5T^a\tilde{G}\psi(x)$, which can be
easily found having the correct quantum numbers of the hybrid and
was also used in Ref.\cite{J. Gov 2}, where, however, the tri-gluon
condensate contribution was not taken into account and the plane
wave method was used. In our work, we keep the operator product
expansion (OPE) to dimension six, the dimension of the tri-gluon
condensate, and take the widespreadly used Fixed-Point gauge
technique. Our numerical result indicates that the contribution of
the tri-gluon condensate is not negligible and even somehow
important for the estimation of the charmonium hybrid mass. In a
similar work done by Kisslinger, Parno, and Riordan \cite{L. S.
Kis}, the current $J_H^\mu(x)=\bar{\Psi}C\gamma_\nu
G^{\mu\nu}\Psi(x)$ was employed to calculate the $1^{--}$ charmonium
hybrid, which we think is improper due to the incompatible quantum
number with the concerned hybrid, and they obtained a quite low mass
of 3.66 GeV.

The rest of this paper is organized as follows. In Sec.II we derive
the formulas of the correlation function $\Pi_{\mu\nu}$ in terms of
the QCD Sum Rules with the interpolating current for
$J^{PC}=1^{--}$. In Sec III, our numerical evaluations and relevant
figures are presented. Section V is remained to summary and
concluding remarks.

\section{Formalism}

In the framework of QCD Sum Rules, the starting point is properly
constructing the correlation function, i.e.,
\begin{eqnarray}
  \Pi_{\mu\nu}(q)&=&i\int d^4x e^{iq\cdot x}
  \langle0|T\{J_\mu(x)J_\nu^\dagger(0)\}|0\rangle\;.
  \label{correlator}
\end{eqnarray}
Here, the interpolating current $J_\mu$ for the heavy hybrid with
quantum number $J^{PC}=1^{--}$ is chosen to be
\begin{eqnarray}
  J_\mu(x)=g_s\bar{\psi}^a(x)\gamma^\nu\gamma_5\frac{\lambda^n_{ab}}{2}
  \tilde{G}^n_{\mu\nu}(x)\psi^b(x)\;,
\end{eqnarray}
where $g_s$ is the strong coupling constant, $a$ and $b$ are color
indices, $\lambda^n$ is the color matrices, and
$\tilde{G}^n_{\mu\nu}(x)=\epsilon_{\mu\nu\alpha\beta}G^{n,\alpha\beta}(x)/2$
is the dual field strength of $G^n_{\mu\nu}(x)$. Generally, the
two-point function $\Pi_{\mu\nu}$ may contain two distinct parts,
the vector part $\Pi_V(q^2)$ and the scalar part $\Pi_S(q^2)$ which
represent the contributions of the correlation function to the
vector channel $J^{PC}=1^{--}$ and scalar channel $J^{PC}=0^{+-}$,
respectively. They can be explicitly expressed as
\begin{eqnarray}
  \Pi_{\mu\nu}(q)&=&(\frac{q_\mu q_\nu}{q^2}
  -g_{\mu\nu})\Pi_V(q^2)+\frac{q_\mu
  q_\nu}{q^2}\Pi_S(q^2)\;.
\end{eqnarray}
Since the main task of our work is to study the mass of the vector
heavy hybrid with the quantum number $1^{--}$, following we only
analyze the vector part $\Pi_V(q^2)$.

\begin{figure}[h]
  \begin{center}
  \includegraphics[width=8cm]{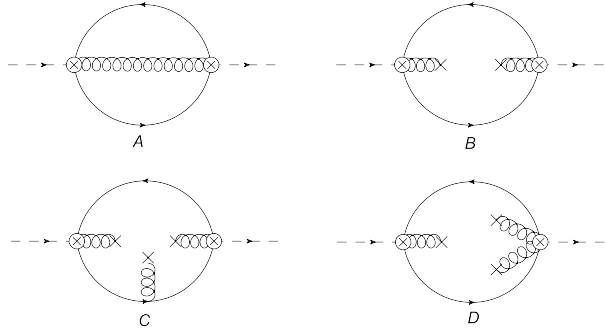}\\
  \caption{The typical Feynman diagrams for
  the calculation of heavy
  vector hybrid masss. Here, the permutation diagrams are implied.
  Diagram (A) represents for the contribution
  of unit operator; (B) for the
  contribution of two-gluon condensate;
  (C) and (D) for the contribution
  of three-gluon condensate.}\label{1}
  \end{center}\label{fig 1}
\end{figure}

By the operator product expansion (OPE), the correlation function
$\Pi_V(q^2)$ can be written as
\begin{eqnarray}
  \Pi_V(q^2)=\Pi^{\text{pert}}(q^2)+\Pi_i^{\text{cond}}(q^2)\;,
\end{eqnarray}
where to satisfy the necessity of our calculation,
$\Pi^{\text{pert}}(q^2)$ is obtained by taking the imaginary part of
the Feynman diagram A, and $\Pi_i^{\text{cond}}(q^2)$ represents the
contributions coming from all possible condensates. In this work, we
consider the condensates up to dimension six.

First, we calculate the imaginary part, the absorptive part, of the
Feynman diagrams which represents the perturbative contribution to
the correlator, and the result reads
\begin{eqnarray}\label{}
 \rho^{\text{pert}}(t) &=&-\frac{\alpha_s m_Q^6}{720\pi^2\sqrt{1-t}\
  t^3}
   \Big[-15t^5+185t^4-778t^3-496t^2+1296t-192\nonumber\\
 & &+15t^2\sqrt{1-t}(t^3-12t^2+48t-128)\log\frac{\sqrt{1-t}+1}
  {\sqrt{t}}\Big]\;,
\end{eqnarray}
where, $t=4m_Q^2/s$, and $m_Q$ is the mass of the heavy quark, and
$\rho^{\text{pert}}(t)\equiv {\rm Im}\,\Pi(t)$.

The contributions of non-perturbative condensates from diagrams in
Figure 1 are
\begin{subequations}
\begin{eqnarray}
  \Pi_4^{\text{cond,B}}(q^2)&=&\int_0^1dx \frac{\langle
  g_s^2G^2\rangle}{48\pi^2}\{[8(1-x)xq^2-11m_Q^2]+
  \ln(\Delta)[2(1-x)xq^2-3m_Q^2]\}\; ,\\
  \Pi_6^{\text{cond,C}}(q^2)&=&\int_0^1dx\frac{\langle
  g_s^3G^3\rangle}{192\pi^2}[3x\ln(\Delta)+\frac{2xm_Q^2}{\Delta}+17x]\; ,\\
  \Pi_6^{\text{cond,D}}
  (q^2)&=&\int_0^1dx\frac{\langle g_s^3G^3\rangle}{384\pi^2}
  \{2x(2-3x)\ln(\Delta)-\frac{[2(3-4x)m_Q^2+x(14x^2-27x+13)
  q^2]x}{\Delta}\nonumber\\
  &+&\frac{(x-1)q^2[3xq^2(x-1)^2+(2-3x)m_Q^2]x^2}{\Delta^2}
  +\frac{2(5-24x)x}{3}\}\; .
\end{eqnarray}
\end{subequations}
Here, $\Delta=-(1-x)xq^2+m_Q^2$, and symbols B, C, D represent the
corresponding diagrams, respectively.

In order to eliminate contributions from higher excited and
continuum states, a well-known procedure, the Borel transformation,
is performed to above obtained results, and then we get
\begin{subequations}
\begin{eqnarray}
  \hat{\bf B}[\Pi_4^{\text{cond,B}}(q^2)]&=&\int_0^1d\omega\int_0^1dx
  \frac{\langle g_s^2G^2\rangle m_Q^4}{48\pi^2(1-x)x\omega^3}
  e^{-\frac{m_Q^2}{(1-x)x\omega M_B^2}}(3\omega-2)\;,\\
  \hat{\bf B}[\Pi_6^{\text{cond,C}}(q^2)])&=&\int_0^1d\omega
  \int_0^1dx\frac{\langle g_s^3G^3\rangle m_Q^2}{192\pi^2(1-x)}
  \bigg(2e^{-\frac{m_Q^2}{(1-x)xM_B^2}}-3\frac{1}{\omega^2}
  e^{-\frac{m_Q^2}{(1-x)x\omega M_B^2}}\bigg)\;,\\
  \hat{\bf B}[\Pi_6^{\text{cond,D}}(q^2)]&=&\int_0^1 d\omega
  \int_0^1 dx\frac{\langle g_sG^3\rangle m_Q^2}{384M_B^2\pi^2
  (x-1)^2x\omega^2}\bigg\{-M_B^2(x-1)x\big[e^{-\frac{m_Q^2}{M_B^2(1-x)x\omega}}
  (6x-4)\nonumber\\
  &&+e^{-\frac{m_Q^2}{M_B^2(1-x)x}}(13x-11)\omega^2\big]+e^{-\frac{m_Q^2}{M_B^2(1-x)x}}
  (6x-5)\omega^2m_Q^2\bigg\}\;,
\end{eqnarray}
\end{subequations}
where,  $M_B$ is the Borel parameter.

Suppose the existence of the quark-hadron duality, the resultant sum
rule for the mass of the vector heavy hybrid reads
\begin{eqnarray}
  m_H = \sqrt{-\frac{R_1}{R_0}}\label{mass}
\end{eqnarray}
  with
\begin{eqnarray}
  R_0 = \frac{1}{\pi}\int^{s_0}_{4m_Q^2}
  \rho^{\text{pert}}(s)e^{-s/M_B^2}& + & \hat{\bf B}(\Pi_4^{\text{cond,B}})
  +\hat{\bf B}(\Pi_6^{\text{cond,C}})+\hat{\bf B}(\Pi_6^{\text{cond,D}})\;,\\
  R_1 & = &\frac{\partial}{\partial{M_B^{-2}}}{R_0}\;.
\end{eqnarray}
Here, $s_0$ is the threshold cutoff introduced to remove the
contribution of the higher excited and continuum states \cite{P.
Col}.

\section{Numerical Analysis}

For numerical calculation, the leading order strong coupling
constant
\begin{equation}
\alpha_s(M_B^2)=\frac{4\pi}{(11-\frac{2}{3}n_f)
\ln(\frac{M_B^2}{\Lambda_{\text{QCD}}^2})}
\end{equation}
is adopted with $\Lambda_{\text{QCD}}=0.220\; \text{GeV}$ and $n_f$
being the number of active quarks. Of the two- and three-gluon
condensates we take both the prevailing values \cite{P. Col}
\begin{eqnarray}
  \langle \alpha_s G^2\rangle=0.038\pm0.011\; \text{GeV}^4\;, \;
  \langle g_s^3G^3\rangle=0.045\; \text{GeV}^6\;\label{condensates1}
\end{eqnarray}
and also recently obtained ones \cite{Narison:2011rn}, i.e.,
\begin{eqnarray}
  \langle \alpha_s G^2\rangle=0.075\pm0.020\; \text{GeV}^4\;, \;
  \langle g_s^3G^3\rangle=(8.3\pm1.0)\text{GeV}^2\;\langle \alpha_sG^2\rangle\;
  \label{condensates2}
\end{eqnarray}
into account. The heavy quark masses are taken to be:
\begin{eqnarray}
  m_{c}=1.26\sim1.47\; \text{GeV}, \ m_{b}=4.22\sim4.72\;
  \text{GeV}\;.
\end{eqnarray}
Here, the masses span from the running masses in $\overline{MS}$
scheme to the on-shell masses of QCD Sum Rules
\cite{Narison:2010py}.

For the selection of an appropriate Borel parameter $M_B^2$, we
adopt the criteria proposed in Refs.\cite{M. A. Shi, L. J. Rei, D.
S. Du}. Defining $m_H(M_B^2)$ in Eq.(\ref{mass}) to be
$f_{\text{thcorr}}(M^2)$ while the continuum contribution is absent,
i.e., ($s_0=\infty$), and $m_H(M_B^2)$ to be
$m_{\text{H,nopower}}(M_B^2)$ in case of no power corrections. Then
we get two discrimination functions $f_{\text{cont}}(M_B^2)$ and
$f_{\text{nopower}}(M_B^2)$, satisfying
\begin{eqnarray}
  f_{\text{cont}}(M_B^2)&=&\frac{m_{H}
  (M_B^2)}{f_{\text{thcorr}}(M_B^2)}\;,\\
  f_{\text{nopower}}(M_B^2)&=&\frac{m_{H}
  (M_B^2)}{m_{\text{H,nopower}}(M_B^2)}\;.
\end{eqnarray}
According to the sum rule criteria, the mass function obtained in
Eq.(\ref{mass}) is valid only in the situation of $M_B^2$ being
neither too small nor too large. In case $M_B^2$ is very small, the
omitted terms of high-dimensional condensates, which are
proportional to high powers of $1/M_B^2$, may become too important
to be neglected. To get a reliable prediction of the hybrid mass in
QCD Sum Rules, $f_{\text{nopower}}(M_B^2)$ is required to be less
than 10\% from unit, which ensures the contributions from the
non-pertubative condensates to be much less than what from the
perturbative term. On the other hand, a very large $M_B^2$ implies
the invalidation of quark-hadron duality approximation. Normally,
the $f_{\text{cont}}(M_B^2)$ is required to be 70\% more to suppress
the contributions of higher resonances and continuum.

\begin{figure}[h]
  \begin{center}
  \includegraphics[width=16cm]{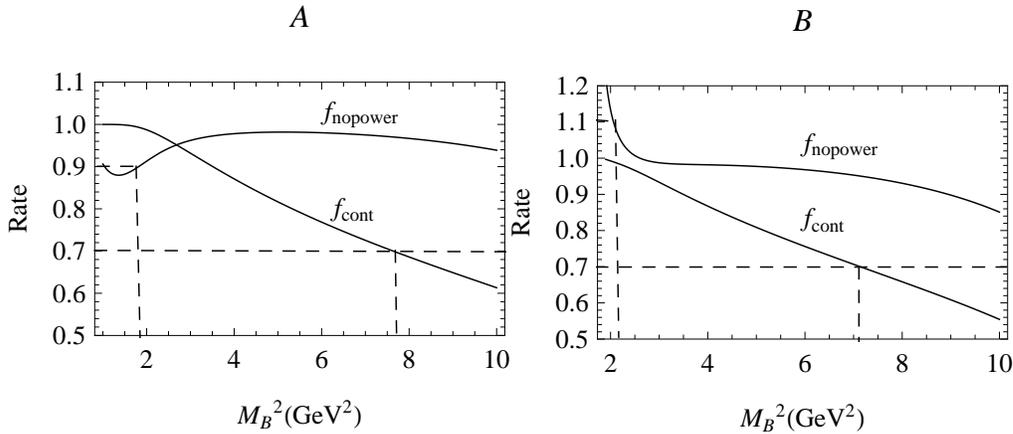}\\
  \caption{The curves of $f_{\text{nopower}}$ and
  $f_{\text{cont}}$ for charmonium hybrid versus Borel
  parameter $M_B^2$ while the continuum threshold cutoff $s_0=26\;
  \text{GeV}^2.$ Figures A and B correspond to the gluon
  condensates from \cite{P. Col} and \cite{Narison:2011rn},
  i.e. Eqs.(\ref{condensates1}) and (\ref{condensates2}),
  respectively.} \label{rate-cc}
  \end{center}
\end{figure}

For charmonium hybrid, to find the reliable sum rule for hybrid mass
according to the aforementioned criteria, in Fig. \ref{rate-cc}, we
draw the curves of $f_{\text{nopower}}$ and $f_{\text{cont}}$ as
functions of $M_B^2$. The figure indicates that $f_{\text{nopower}}$
and $f_{\text{cont}}$ may satisfy the above mentioned requirements,
i.e., the contribution from pole terms is more than 70\% and the
contribution from condensates is less than 10\%, while $M_B^2$ lies
in between $1.80\;$ to $7.80\;\text{GeV}^2$ for
Eq.(\ref{condensates1}) and between $2.10\;$ to $7.20\;\text{GeV}^2$
for Eq.(\ref{condensates2}), respectively.

\begin{figure}[h]
  \begin{center}
  \includegraphics[width=16cm]{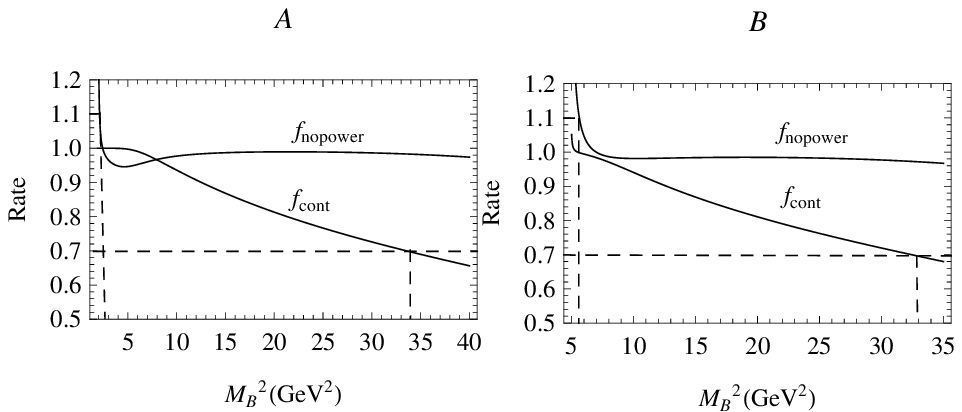}\\
  \caption{The curves of $f_{\text{nopower}}$
  and $f_{\text{cont}}$ for bottomnium hybrid versus Borel
  parameter $M_B^2$ while the continuum threshold cutoff $s_0=130\;
  \text{GeV}^2.$ Figures A and B correspond to the gluon
  condensates from \cite{P. Col} and \cite{Narison:2011rn},
  i.e. Eqs.(\ref{condensates1}) and (\ref{condensates2}),
  respectively.} \label{rate-bb}
  \end{center}
\end{figure}

For bottomnium hybrid, as shown in figure 3 the reliable range for
$M_B^2$ lies in between $3.00\;$ to $34.00\;\text{GeV}^2$ for
Eq.(\ref{condensates1}) and between $5.00\;$ to
$33.00\;\text{GeV}^2$ for Eq.(\ref{condensates2}), respectively.

With the above requirements in establishing the sum rule, to
calculate the physical quantities in terms of QCD Sum Rules, one
needs to find an optimal window for Borel parameter $M_B^2$ and
threshold parameter $s_0$. Within this window, the physical
quantities, here the hybrid mass, are maximally independent of the
Borel parameter.

\begin{figure}[h]
  \begin{center}
  \includegraphics[width=16cm]{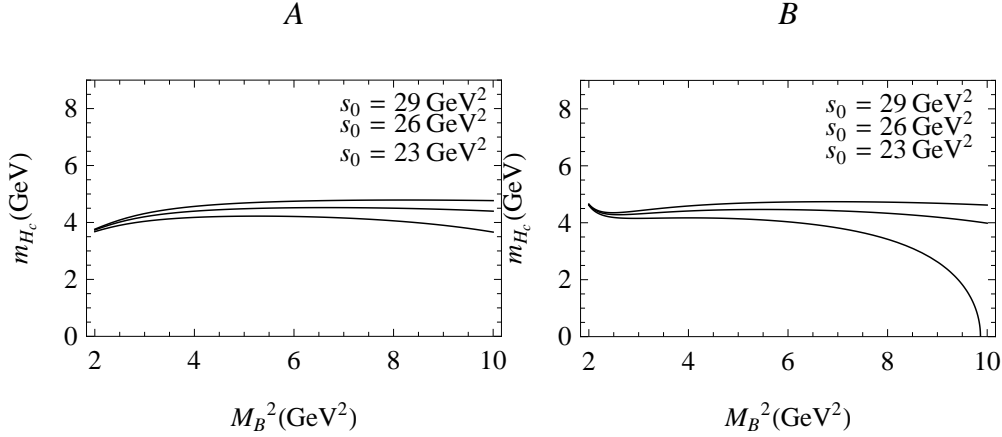}\\
  \caption{The dependence of $1^{--}$ charmonium hybrid sum-rule mass
  on the Borel parameter
  $M_B^2$ with different continuum threshold cutoffs $s_0$, that is
  $29\; \text{GeV}^2$, $26\; \text{GeV}^2$ and $23\; \text{GeV}^2$
  from up to down in the figure. Figures A and B correspond to the gluon
  condensates from \cite{P. Col} and \cite{Narison:2011rn},
  i.e. Eqs.(\ref{condensates1}) and (\ref{condensates2}),
  respectively.}\label{mass-cc}
\end{center}
\end{figure}

In figure \ref{mass-cc}-A, we draw lines for different threshold
parameter $s_0$, that is 29, 26, and 23$\; \text{GeV}^2$ while Borel
parameter varying from $1.80$ to $7.80$$\; \text{GeV}^2$. The figure
evidently indicates that there exists a window for the Borel
parameter between $4.20\; \text{GeV}^2$ and $7.80\; \text{GeV}^2$,
in which the evaluated hybrid mass is mostly independent of $M_B^2$,
especially in case of $s_0 = 26\; \text{GeV}^2$. Namely, it is
proper to select the threshold parameter to be 26$\; \text{GeV}^2$,
which hints that the mass of the first excited state is above $5.10$
GeV. In figure \ref{mass-cc}-B, the corresponding parameters are
$s_0=29, 26, 23 $$\; \text{GeV}^2$,
$2.50\text{GeV}^2<M_B^2<7.20\text{GeV}^2$, and the proper threshold
parameter is also 26$\; \text{GeV}^2$.

Considering the uncertainties remain in the input parameters, the
quark mass, condensates, the Borel parameter $M_B^2$ and continuum
threshold $s_0$, we obtain the charmonium hybrid mass to be
\begin{eqnarray}
 m_{H_c} = 4.52^{+0.27}_{-0.38}\;\text{GeV}\;.\label{massc}
\end{eqnarray}
Here, the charm quark mass $m_c$ goes from $1.26
\sim1.47\;\text{GeV}$; the condensates take the magnitudes of
Eq.(\ref{condensates1}); the Borel parameter $M_B^2$ varies from
$4.20\;\text{GeV}^2$ to $7.80\;\text{GeV}^2$; and the continuum
threshold $s_0$ changes from $23\;\text{GeV}^2$ to
$29\;\text{GeV}^2$. The central value of $m_{H_c}$ in
Eq.(\ref{massc}) is reached by taking the central values of the
quark mass and condensates, while setting $s_0=26\; \text{GeV}^2$
and $M_B^2=6.00\; \text{GeV}^2$.

For the gluon condensates taken as Eq.(\ref{condensates2}), we
obtain the charmonium hybrid mass to be $m_{H_c} =
4.45^{+0.28}_{-0.32}\;\text{GeV}$ with $s_0=26\text{GeV}^2$ and
$M_B^2=6.00\text{GeV}^2$, which is slightly higher than
(\ref{massc}).

In bottomnium sector, by the same procedure, but with inputs
$s_0=130\text{GeV}^2$ and $M_B^2=20\text{GeV}^2$ as shown in figure
5, we readily obtain the $1^{--}$ bottomonium hybrid mass, that is
$m_{H_b}=10.81^{+0.23}_{-0.24}\;\text{GeV}$ and
$10.70^{+0.45}_{-0.46}\text{GeV}$ for condensates from \cite{P. Col}
and \cite{Narison:2011rn}(Eqs.(\ref{condensates1}) and
(\ref{condensates2})), respectively.

\begin{figure}[h]
  \begin{center}
  \includegraphics[width=16cm]{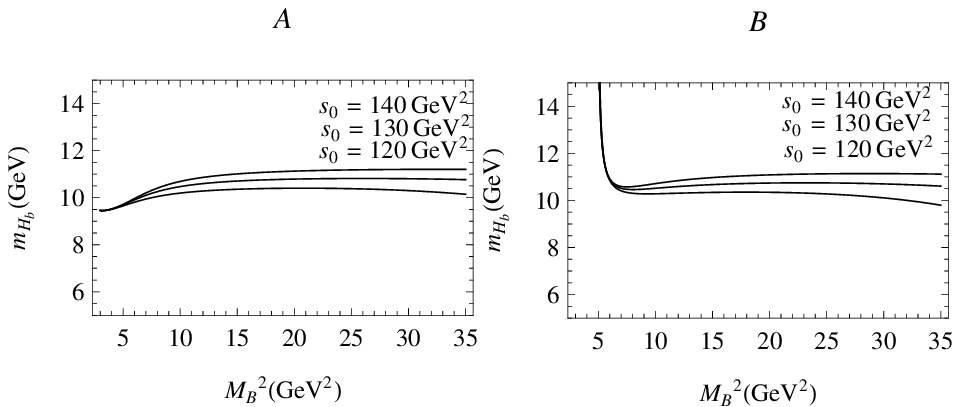}\\
  \caption{The dependence of $1^{--}$ bottomnium hybrid sum-rule mass
  on the Borel parameter
  $M_B^2$ with different continuum threshold cutoffs $s_0$, that is
  $140\; \text{GeV}^2$, $130\; \text{GeV}^2$ and $120\; \text{GeV}^2$
  from up to down in the figure. Figures A and B correspond to the gluon
  condensates from \cite{P. Col} and \cite{Narison:2011rn},
  i.e. Eqs.(\ref{condensates1}) and (\ref{condensates2}),
  respectively.}\label{3}
\end{center}
\end{figure}

\section{Summary and Conclusions}

In this work we recalculate the $1^{--}$ heavy quarkonium masses in
the framework of QCD Sum Rules. The central part of this calculation
relys on the evaluation of the Wilson coefficients of the operators
for the two point correlation function constructed with a suitable
hybrid current. In former studies \cite{J. Gov 2,L. S. Kis}, people
also fought the same target but with some differences from this
work. In previous works, operator expansion only up to dimension
four, but in our study the dimension six operators are taken into
account. And our results indicate that the dimension six operator,
the three-gluon condensate, is important in attaining a wide stable
plateau and hence a stable sum rule, which makes the predictions for
the masses of $c\bar{c} G$ and $b\bar{b} G$ hybrid states more
reliable. We find the interpolating current used in Ref.\cite{L. S.
Kis} is improper, and our procedure in establishing the sum rule
differs from what performed in Ref.\cite{J. Gov 2}. In our
calculation, the central values of charmonium- and
bottomonium-hybrid masses are 4.52 GeV and 10.81 GeV for condensates
in (\ref{condensates1}) and 4.45 GeV and 10.70 GeV for condensates
in (\ref{condensates2}), respectively.

Considering of the $Y$ states in charmonium region which are
observed recently at the B-factories, since our predicted mass of
the charmonium hybrid resides between $Y(4360)$ and $Y(4660)$, we
are tempted to conclude that neither of the $Y(4260)$, $Y(4360)$ and
$Y(4660)$ states can attribute to pure charmonium hybrid state. If
we take the errors into consideration seriously, the masses of
$Y(4360)$ and $Y(4660)$ are closer to our estimation, thus might be
candidates. However, at present stage it is still hard to draw a
definite conclusion. As a matter of fact, by the discussion given in
the introduction, a pure hybrid or glueball might not independently
exist. The observed structures could be admixtures of relevant
hybrid with glueballs and regular quarkonia with the same quantum
numbers \cite{Close,Hexg}. Therefore, in this sense it is
understandable why the theoretical hybrid mass does not coincide
with any of the resonances observed in experiments. To clarify this
issue, i.e., to calculate the mixing angles among states with the
same quantum numbers, we need a bigger database on the exotic
states, which might be available at the LHCb and planned
Super-Flavor factory.

Finally, we have also estimated the mass of $1^{--}$ bottomonium
hybrid state with quite small uncertainties, the $b\bar bG$, which
is compatible with the recent BELLE observation of the exotic
$Y_b(10890)$ \cite{K.F.Chen}, and may confront to the LHCb data in
near future.

\vspace{.7cm} {\bf Acknowledgments} \vspace{.3cm}

This work was supported in part by the National Natural Science
Foundation of China, by the CAS Key Projects KJCX2-yw-N29 and
H92A0200S2, and by the special foundation for Ph.D program of the
Ministry of Education of China. L.T. would like to thank X. H. Yuan
for helpful discussion.


\end{document}